\documentclass[conference]{IEEEtran}
\IEEEoverridecommandlockouts
\usepackage{cite}
\usepackage{amsmath,amssymb,amsfonts}
\usepackage{algorithm}
\usepackage{algpseudocode}
\usepackage{graphicx}
\usepackage{textcomp}
\usepackage{xcolor}
\usepackage{subfigure}
\usepackage{geometry}
\def\BibTeX{{\rm B\kern-.05em{\sc i\kern-.025em b}\kern-.08em
    T\kern-.1667em\lower.7ex\hbox{E}\kern-.125emX}}
\begin{document}
\newgeometry{left=0.75in,right=0.75in,top=1in,bottom=0.75in}
\title{K-BMPC: Derivative-based Koopman Bilinear Model Predictive Control for Tractor-Trailer Trajectory Tracking with Unknown Parameters
\thanks{
All authors are with Department of Automation, Institute of Medical Robotics, Shanghai Jiao Tong University, and Key Laboratory of System Control and Information Processing, Ministry of Education of China, and Shanghai Engineering Research Center of Intelligent Control and Management, Shanghai 200240, China. Han Zhang is the corresponding author (email: zhanghan\_tc@sjtu.edu.cn).

This work was supported by the National Key R\&D Program of China under Grant 2022YFC3601403, and the ZF (China) Investment Co., Ltd}
}
\author{
\IEEEauthorblockN{Zehao Wang, Han Zhang, and Jingchuan Wang}
}
\maketitle

\begin{abstract}
Nonlinear dynamics bring difficulties to controller design for control-affine systems such as tractor-trailer vehicles, especially when the parameters in the dynamics are unknown. To address this constraint, we propose a derivative-based lifting function construction method, show that the corresponding infinite dimensional Koopman bilinear model over the lifting function is equivalent to the original control-affine system.  Further, we analyze the propagation and bounds of state prediction errors caused by the truncation in derivative order. 
The identified finite dimensional Koopman bilinear model would serve as predictive model in the next step. Koopman Bilinear Model Predictive control (K-BMPC) is proposed to solve the trajectory tracking problem. We linearize the bilinear model around the estimation of the lifted state and control input. Then the bilinear Model Predictive Control problem is approximated by a quadratic programming problem. Further, the estimation is updated at each iteration until the convergence is reached.
Moreover, we implement our algorithm on a tractor-trailer system, taking into account the longitudinal and side slip effects. The open-loop simulation shows the proposed Koopman bilinear model captures the dynamics with unknown parameters and has good prediction performance. Closed-loop tracking results show the proposed K-BMPC exhibits elevated tracking precision with the commendable computational efficiency. The experimental results demonstrate the feasibility of K-BMPC.
\end{abstract}

\section{Introduction}
Tractor-trailer vehicles are widely used nowadays, particularly in fields such as agriculture and logistics due to their large cargo capacity and high transport efficiency. 
Despite their numerous benefits, achieving high-accuracy tractor-trailer tracking control is challenging, particularly for optimization-based trajectory planning methods\cite{Li 2020, Li 2021}. These methods discretize the kinematic constraints with a large sampling period to reduce the number of optimization variables. Nevertheless, the outcome of such methods would violate the dynamics of tractor-trailer to great extent. Therefore, a controller is needed for tractor-trailer vehicles to follow the trajectories with low tracking errors and high computational efficiency. Model Predictive Control (MPC) presents an attractive approach to trajectory tracking control, due to its adaptability to performance metrics and constraints\cite{Kayacan 2018}. However, MPC problem becomes difficult to solve in real time because of the nonlinear terms in model dynamics and long prediction horizons. Compared to nonlinear models, locally linearized models carry advantages in computational efficiency. However, their accuracy declines when the vehicle states move away from the point of linearization\cite{Gros 2020}.

In addition to the locally linearized models, Koopman operator has been gaining attention for its ability to predict the flow of nonlinear dynamics using an infinite-dimensional linear model\cite{Koopman 1931}. Extended Dynamic Mode Decomposition (EDMD) and is a data-driven tool to identify finite dimensional approximations of the Koopman operator and hence applied to approximate a variety of nonlinear dynamics \cite{Korda 2018, Bruder 2020, Gupta 2022, Mamakoukas 2021}. However, the lifting functions used to construct EDMD-based Koopman models generally depend on the expert selections\cite{Cibulka 2019}. This means a lot of tuning in practice. To this end, deep neural networks are employed to overcome the difficulties in lifting function construction\cite{Han 2020, Shi 2022, Xiao 2022}, however the interpretability of the model is limited.

Prior work on Koopman-based MPC design primarily focuses on combining linear MPC with the linear lifted models to increase the computational speed, however the accuracy of the lifted linear model is not guaranteed when the original states and controls are coupled. 
Koopman bilinear models are considered to balance the accuracy and computational speed, and the characteristics such as the bilinearizability and reachability are proved in \cite{Goswami 2021}. However, the bilinear term brings difficulties in controller design. To address the difficulties, a few works try to linearize the bilinear models based on the current state of the system\cite{Bruder 2021}, but the linearization suffers from the disadvantage of local linearization. Folkestad et al. solve Koopman nonlinear MPC using sequential quadratic programming, however the Hessian of Lagrangian can not be computed directly from the bilinear structure\cite{Folkestad 2021, Folkestad 2022}.

The challenge in tractor-trailer trajectory tracking using MPC is the nonlinearity in the dynamics. To address the challenge, we generalize derivative-based Koopman operators \cite{Mamakoukas 2021} to Koopman bilinear models, transform the tractor-trailer dynamic into a bilinear model. Then, we propose an iterative strategy to solve the Koopman bilinear MPC problems. Simulation and experimental results demonstrate strengths of our method. Our contributions are twofold. 
\restoregeometry
\begin{itemize}
    \item We propose a lifting function construction method based on the derivatives of the dynamics, show that the corresponding infinite dimensional Koopman bilinear model is equivalent to the original control-affine system.
    Moreover, under the assumption that the derivative order is truncated, we analyze the state prediction error propagation and its bounds. 
\item 
    We propose a Koopman bilinear MPC framework (K-BMPC) to solve the bilinear MPC problem using iterative quadratic programming. In K-BMPC, the bilinear model is linearized around the estimation of the state and control input. Then we transform the original MPC problem to a quadratic programming (QP) problem.
\end{itemize}

\section{TRACTOR-TRAILER SYSTEM}
\begin{figure}[htbp]
\centerline{\includegraphics[width=0.6\linewidth]{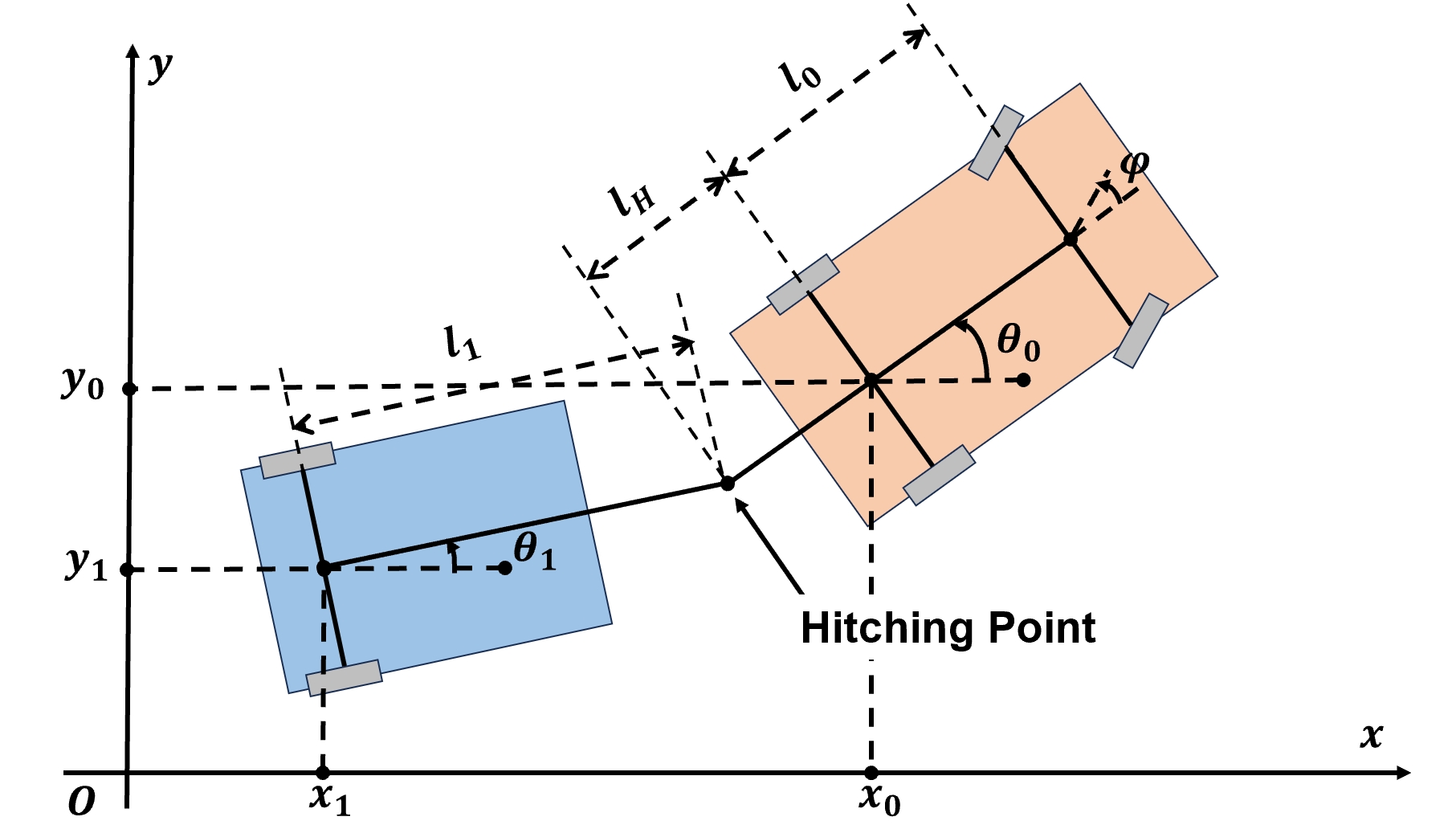}}
\caption{Schematic of a tractor-trailer vehicle system.}
\label{tractor trailer fig}
\end{figure}

In this section, we present the dynamics and constraints of a tractor-trailer vehicle with unknown slip parameters. As shown in Fig.\ref{tractor trailer fig}, the tractor-trailer vehicle system is formed by a tractor towing a trailer. The lengths of the tractor and trailer are denoted by $l_0$ and $l_1$ respectively, and $l_H$ represents the hitching offset distance.

Compared to classical tractor-trailer trajectory tracking control methods \cite{Wu 2017, Ljungqvist 2016}, we take the velocity as a variable rather than a constant value. The longitudinal and side slips are also considered to achieve high accuracy control. Hence, the dynamic model of a tractor-trailer system with unknown slip parameters takes the form 
\begin{equation}
\dfrac{d}{dt}
\begin{bmatrix}
x_0\\
y_0\\
\theta_0\\
\theta_1\\
tan\varphi\\
v\\
\end{bmatrix}
= \begin{bmatrix}
\mu v\cos\theta_0\\
\mu v\sin\theta_0\\
\mu v \tan\kappa\varphi/ l_0\\
\mu v(\sin\delta\theta - \tan \kappa\varphi\cos\delta\theta l_H / l_0)/l_1\\
\omega\\
a\\
\end{bmatrix}.
\label{dynamic model of tractor-trailer}
\end{equation}
The state vector is defined as $x = [x_0, y_0, \theta_0, \theta_1, tan\varphi, v]^T$, where we denote $(x_0, y_0)$ as the position of tractor. $\theta_0$ and $\theta_1$ are defined as the orientation angle of tractor and trailer respectively. $\varphi$ is the front-wheel steering angle and $v$ is the linear velocity. $\delta\theta = \theta_0 - \theta_1$ denotes the jack-knife angle. Further, The control vector is defined as $u = [\omega, a]^T$, where $\omega$ is the derivative of $tan\varphi$ and $a$ is acceleration. $\mu$ and $\kappa$ are unknown traction parameters representing the longitudinal and side slip influence respectively \cite{Kayacan 2018}.  

The aim of tractor-trailer tracking control is to keep the position and orientation of the tractor and trailer close to the reference trajectory.
To this end, we define the output $y = [x_0, y_0, \theta_0, \theta_1, tan\varphi, v, x_1, y_1]^T$ by adding the position of the trailer $(x_1, y_1)$ to the state vector. $(x_1, y_1)$ are computed through the geometrical relationship.
\begin{equation}
\label{trailer position}
\begin{aligned}
x_1 &= x_0 - l_H\cos\theta_0 - l_1\cos\theta_1,\\
y_1 &= y_0 - l_H\sin\theta_0 - l_1\sin\theta_1,\\
\end{aligned}
\end{equation}
the dynamic of $(x_1, y_1)$ is computed through \eqref{dynamic model of tractor-trailer} and \eqref{trailer position}.

In addition to the dynamics, we consider input and output constraints due to the physical limits. The input constraints take the form
\begin{equation}
\label{input constraints}
-u_{max} \leq u \leq u_{max},
\end{equation}
where $u_{max} = [\omega_{max}, a_{max}]^T$ represent maximum steering angle velocity and linear acceleration respectively. The output constraints are defined as
\begin{equation}
\label{output constraints}
\begin{aligned}
-\tan\varphi_{max} \leq &\tan\varphi \leq \tan\varphi_{max}, \\
-v_{max} \leq &v \leq v_{max}, \\
-\delta\theta_{max} \leq & \theta_0  - \theta_1 \leq \delta\theta_{max}, \\
\end{aligned}
\end{equation}
where $\tan\varphi_{max}$ and $v_{max}$ represent maximum steering angle and linear velocity respectively, $\delta\theta_{max}$ is maximum jack-knife angle for collision avoidance between the tractor and trailer.

\section{KOOPMAN THEORY}
In this section, we describe the Koopman operator, its identification methods and  bilinear realization. Moreover, we elaborate the lifting function construction based on the derivatives of the dynamics, analyze the propagation and bounds of  the prediction errors under truncated derivatives. 
\subsection{Koopman operator and EDMD for controlled system}
For a control-affine continuous-time system of the form
\begin{equation}
    \label{nonlinear dynamic}
    \begin{aligned}
    &\dot{x}  = f(x) + \sum_{i=1}^m g_i(x)u_i = F(x,u),y = h(x),\\
    \end{aligned}
\end{equation}
where $x \in \mathbb{R}^{n_x}$ is the state, $u \in \mathbb{R}^m$ is the input and $y \in \mathbb{R}^{n_y}$is the output.
Denote the observable $\phi \in \mathcal{F}: \mathbb{R}^{n_x \times m} \rightarrow \mathbb{R}$, where
$\mathcal{F}$ is an infinite-dimensional function space that is composed of
all square-integrable real-valued functions with compact domain $X \times U \subset \mathbb{R}^{n_x} \times \mathbb{R}^{m}$. The continuous time Koopman operator $\mathcal{K}: \mathcal{F} \rightarrow \mathcal{F}$ describes the evolution of each observable in \eqref{nonlinear dynamic}
\begin{equation}
    \label{Koopman operator}
    {\mathcal{K}\phi}(x,u) = \dfrac{\partial \phi}{\partial x}F(x,u).\nonumber
\end{equation}
Moreover, the corresponding discrete-time Koopman operator $\mathcal{K}_{T_s}$ with sampling time $T_s$ takes the form
\begin{equation}
    \label{ discrete time Koopman operator}
    {\mathcal{K}_{T_s}} = e^{T_s\mathcal{K}}.
    \nonumber
\end{equation}
The dimension of Koopman operator $\mathcal{K}$ is typically infinite, so we construct a finite-dimensional approximation using Extended Dynamic Mode Decomposition (EDMD). 

More specifically, EDMD approximates the discrete-time Koopman operator over a lifting function $\psi: X \times U \rightarrow \mathbb{R}^M$ using experimental data $\{x^{(k)},u^{(k)}\}_{k=1}^K$. 
The approximated $\tilde{\mathcal{K}}_{T_s}$ is the solution to the optimization problem
\begin{equation}
    \label{EDMD koopman operator}
\min\limits_{\tilde{\mathcal{K}}} \sum_{k=1}^{K-1} \Vert \tilde{\mathcal{K}}^T \psi(x^{(k)}, u^{(k)}) - \psi(x^{(k+1)}, u^{(k)})\Vert_2^2.
\end{equation}
\subsection{Koopman Bilinear Model Realization}
We focus on using continuous Koopman bilinear model with the following form to predict the evolution of states in \eqref{nonlinear dynamic}
\begin{equation}
    \label{continuous bilinear dynamic}
    \begin{aligned}
    &\frac{d}{dt}{z}  = \Bar{A}z + \Bar{B}u + \sum_{j=1}^{m} u_{j} \Bar{H}_jz,\\
    &y = Cz,z(0) = \psi_x(x(0)),\\
    \end{aligned}
\end{equation}
where $z \in \mathbb{R}^N (N >> n) $ is the lifted state, $\Bar{A} \in \mathbb{R}^{N \times N}$, $\Bar{B} \in \mathbb{R}^{N \times m} $, $ \Bar{H}_j \in \mathbb{R}^{N \times N}$ and $C \in \mathbb{R}^{n_y \times N}$. The initial condition of lifted state $z(0))$ is given by $z(0) = \psi_x(x(0))$, where 
\begin{equation}
    \psi_x(x) = [h^T(x), \phi_{i}(x)|i=1,\dots, N-n_y]^T.
    \nonumber
\end{equation} 
The first $n_y$ observables are defined as the output vector $y$ for convenience in tracking the trajectory.
In practice, we use the following discrete-time model
\begin{equation}
    \label{bilinear dynamic}
    \begin{aligned}
    &z_{k+1}  = Az_{k} + Bu_{k} + \sum_{j=1}^{m} u_{j,k} H_jz_{k},\\
    &y_k = Cz_k,z_0 = \psi_x(x_0).\\
    \end{aligned}
\end{equation}
To identify matrices $A, B, \{H_j\}$ in \eqref{bilinear dynamic}, we stack $\psi_x(x)$ with the product of $[\psi_x(x),1]^T$ and $u$ to obtain a $(m+1)N+m$ dimensional lifting function
\begin{equation}
\label{lifting function psi bilinear}
\psi(x,u) = [\psi_{x}^T(x), u	\otimes[\psi_{x}(x), 1]^T]^T.
\nonumber
\end{equation}
Then, \eqref{EDMD koopman operator} is solved to compute an approximation of the discrete-time Koopman operator $\tilde{\mathcal{K}}_{T_s}$.
We are able to obtain the matrices $A, B, \{H_j\}$ in \eqref{bilinear dynamic} from $\tilde{\mathcal{K}}_{T_s}$
\begin{equation}
\label{A,B,N in K}
\tilde{\mathcal{K}}_{T_s} = 
\begin{bmatrix}
    A & H_1 & \cdots & H_m & B\\
    \vdots & \vdots & \vdots & \vdots & \vdots\\
\end{bmatrix}^T.
\nonumber
\end{equation}
$C = [I, 0]$ because the first $n_y$ row of lifting function is the output of the system. 

\subsection{Derivative-based lifting function design for control-affine system}

For control-affine system \eqref{nonlinear dynamic}, we denote 
$F_{i}(x,u), f_{i}(x)$, $g_{i}(x)$ and $ h_{i}(x)$ as i-th row of $F(x,u), f(x), g(x)$ and $h(x)$ respectively. More specifically, the $0$-th order derivative is defined as the dynamic itself
\begin{equation}
\label{derivative of F}  
f_i(x)\! + \!\sum_{j=1}^m g_{i,j}(x)u_j \triangleq F_{i,0}^{(0)}(x)\! +\! \sum_{j=1}^m F_{i,j}^{(0)}(x)u_j.
\end{equation}
Then, we define $F_{i}^{(n+1)}(x) = [F_{i,0}^{(n+1)}(x), F_{i,1}^{(n+1)}(x),\cdots]$ from the $n$-th derivatives $F_{i}^{(n)}$
\begin{equation}
\label{derivative of Fn}
\begin{aligned}
\frac{d}{dt} F_{i,l}^{(n)}(x)& = \frac{\partial F_{i,l}^{(n)}}{\partial x}^T[f(x) + \sum_{j=1}^m g_{j}(x)u_j]\\
&= \frac{\partial F_{i,l}^{(n)}}{\partial x}^T f(x) + \sum_{j=1}^m \frac{\partial F_{i,l}^{(n)}}{\partial x}^T g_{j}(x)u_j\\
&\triangleq F_{i,(m+1)l}^{(n+1)}(x) + \sum_{j=1}^m F_{i,(m+1)l+j}^{(n+1)}(x)u_j,\\
\end{aligned}
\end{equation}
where $F_{i,l}^{(n)}$ is the $l$-th element in $F_{i}^{(n)}$, each $F_{i,l}^{(n)}$ will induce $m+1$ elements in $F_{i}^{(n+1)}$.
Then, we define $h_{i}^{(n)}(x)=[h_{i,0}^{(n)}(x), h_{i,1}^{(n)}(x), \cdots]$ in a similar way to \eqref{derivative of F}, \eqref{derivative of Fn}.

Based on the derivatives of dynamics, we construct an infinite-dimensional lifting function
\begin{equation}
\label{infinite lifting function psi}
\begin{aligned}
&\psi_x(x) = \left [ h^{(n)}(x), F^{(n)}(x), n \geq 0 \right ]^T,\\
\end{aligned}
\end{equation}
where the $F^{(n)}(x) \triangleq [F_{1}^{(n)}(x),\dots,F_{n_x}^{(n)}(x)]$, $h^{(n)}(x)\triangleq [h_{1}^{(n)}(x),\dots,h_{n_y}^{(n)}(x)]$. According to the necessary and sufficient conditions for bilinear realizations \cite{Bruder 2021}, the infinite dimensional Koopman realization \eqref{continuous bilinear dynamic} over \eqref{infinite lifting function psi} is bilinear.  

In practice, we truncate the lifting function to obtain a finite dimensional bilinear model by limiting the order of derivatives less than or equal to a pre-defined degree $\rho$
\begin{equation}
\label{finite lifting function psi}
\psi_x(x) = \left [ h^{(n)}(x), F^{(n)}(x), 0 \leq n \leq \rho \right ]^T.
\end{equation}
The truncation in the derivative order would bring errors to the state and output estimation.
The corresponding truncated bilinear model is of the form
\begin{equation}
\label{estimation x_dot}
\frac{d}{dt}\!
\begin{bmatrix}
\tilde{x}_i\\
F_{i,0}^{(0)}(\tilde{x})\\
\vdots\\
F_{i,j}^{(0)}(\tilde{x})\\
\vdots\\
F_{i,0}^{(\rho)}(\tilde{x})\\
\vdots\\
\end{bmatrix} \! =\!
\begin{bmatrix}
F_{i,0}^{(0)}(\tilde{x})\! +\! \sum_{j=1}^m F_{i,j}^{(0)}(\tilde{x})u_j\\
F_{i,0}^{(1)}(\tilde{x})\! +\! \sum_{j=1}^m F_{i,j}^{(1)}(\tilde{x})u_j\\
\vdots\\
F_{i,m+1}^{(1)}(\tilde{x})\! +\! \sum_{j=1}^m F_{i,m+1+j}^{(1)}(\tilde{x})u_j\\
\vdots\\
0\\
\vdots\\
\end{bmatrix} \!,
\end{equation}
where $\tilde{x}$ is the estimation of $x$. Note that $F_{i,0}^{(n)}(\tilde{x})(n > \rho)$ is set to $0$. To evaluate the prediction error between the truncated bilinear model and the original control-affine model, we denote 
\begin{equation}
\label{F n}
\begin{aligned}
& F_i^{(0)}(x,u) = F_{i,0}^{(0)}(x) + \sum_{j=1}^m F_{i,j}^{(0)}(x)u_j,\\
& F_i^{(n+1)}(x,u) = \frac{d}{dt}F_i^{(n)}(x,u).\\
\nonumber
\end{aligned}
\end{equation}
The propagation from $\tilde{x}_{i,k}$ to $\tilde{x}_{i,k+1}$ under truncated bilinear model \eqref{estimation x_dot} takes the form
\begin{equation}
\label{estimation x}
\tilde{x}_{i,k+1} = \tilde{x}_{i,k} + F_{i}^{(0)}(\tilde{x}_k,u_k)T_s + \cdots + F_{i}^{(\rho)}(\tilde{x}_k,u_k)\frac{T_s^\rho}{(\rho)!}.
\nonumber
\end{equation}
The propagation of state estimation error $e_{i,k} = x_{i,k} - \tilde{x}_{i,k}$ is   
\begin{equation}
\label{error ek}
e_{i,k+1}\! =\! e_{i,k}\! + \!e_{i,k}^{(1)}T_s\! + \cdots + \! e_{i,k}^{(\rho+1)}\frac{T_s^\rho}{\rho !} \! + \! F_{i}^{(\rho\!+\!1)} (x_t,u_t) \frac{T_s^{\rho+1}}{(\rho\!+\!1)!},
\nonumber
\end{equation}
where $e_{i,k}^{(j+1)} = F_i^{(j)}(x_k,u_k) - F_i^{(j)}(\tilde{x}_k,u_k)$ and $t \in [t_k, t_{k+1}]$. The bounds and propagation of errors are written in a similar way to \cite{Mamakoukas 2021}
\begin{equation}
\label{propagation error ek}
e_{i,k} = \sum_{j=1}^{k-1}\sum_{p=1}^{\rho} e_{i,j}^{(p)}\frac{T_s^p}{p !} + \sum_{j=0}^{k-1} F_{i,j,j+1}^{(\rho+1)} (x,u)\frac{T_s^{\rho+1}}{(\rho+1)!},
\nonumber
\end{equation}
\begin{equation}
\label{max ek}
|e_{i,k}| \leq \frac{(kT_s)^{\rho+1}}{(\rho+1)!} F_{i,max}^{(\rho)} (x,u), 
\nonumber
\end{equation}
where $F_{i,max}^{(\rho+1)} (x,u)$ is the maximum of $F_{i}^{(\rho+1)} (x,u)$.
\section{BILINEAR MODEL PREDICTIVE CONTROL}
\subsection{Formulation of BMPC}
Given a reference trajectory $\{y_k^{r}\}$. We formulate BMPC problem with bilinear models and linear constraints as 
\begin{equation}
    \label{BMPC}
    \begin{aligned}
    \min\limits_{{\{u_k\}} \atop {{\{z_k\}}\atop {\{y_k\}}}}
    &\sum_{k=0}^{N_p} (y_{k} -y_{k}^{r})^T Q_{k}(y_{k} - y_{k}^{r}) + \sum_{k=0}^{N_p-1}u_k^TRu_k\\
    s.t.& z_{k+1}\! = \!Az_{k}\! + \!Bu_{k} \! + \!\sum_{j=1}^{m} u_{j,k} H_jz_{k}, k=0,\dots,N_p-1\\
    &y_k = Cz_k, z_0 = \psi_x(x_0), \qquad \qquad k=0,\dots,N_p\\
    &E_k y_k \! + \!F_k u_k\! \leq\! l_k, E_{N_p} y_{N_p}\! \leq\! l_{N_p}, \ k=0,\dots,N_p-1\\
    \end{aligned}  
\end{equation}
where $N_p$ is prediction horizon, $Q_k \in \mathbb{R}^{N_y \times N_y}$ and $ R_k \in \mathbb{R}^{m \times m}$ are positive semidefinite, $E_k \in \mathbb{R}^{n_c \times N_y}$, $ F_k \in \mathbb{R}^{n_c \times m}$ and $ l_k \in \mathbb{R}^{n_c}$ define the input and output constraints in \eqref{input constraints} and \eqref{output constraints} for brevity, where $n_c$ is the number of constraints. Due to the bilinear term in predictive model, there is no explicit solution to the problem, hence numerical scheme is needed. We linearize the bilinear term around the estimation of $z_k$ and $u_k$
\begin{equation}
    \label{linearize bilinear dynamic}
    z_{k+1}  = \hat{A}_kz_{k} + \hat{B}_ku_{k}- \sum_{j=1}^{m} \hat{u}_{j,k} H_j\hat{z}_{k},
\end{equation}
where $\hat{z}_k$ and $\hat{u}_k$ are estimations of $z_k$ and $u_k$, and
\begin{equation}
    \label{A_hat, B_hat}
    \hat{A}_{k}  = (A+\sum_{j=1}^{m} \hat{u}_{j,k} H_j), \hat{B}_k = (B+[H_1\hat{z}_{k},\dots,H_m\hat{z}_{k}]).
\end{equation}
\subsection{Iterative quadratic programming}

We linearize the bilinear term through \eqref{linearize bilinear dynamic}, however the accuracy of the linearization is determined by the accuracy of $\{\hat{z}_k\}$. Thus, we design an iterative quadratic programming method to make $\{\hat{z}_k\}$ approach $\{z_k\}$ iteratively. At $i$-th iteration time, by replacing $\{z_k\}$, $\{u_k\}$ and $\{y_k\}$ in \eqref{BMPC} with 
$\hat{z}_{i,k} + \Delta z_k$, $\hat{u}_{i,k} + \Delta u_k$ and $\hat{y}_{i,k} + \Delta y_k$ then applying \eqref{linearize bilinear dynamic}, the BMPC problem is converted to a linearized MPC problem
\begin{equation}
    \label{I-MPC}
    \begin{aligned}
    \min\limits_{{\{\Delta u_k\}} \atop {{\{\Delta z_k\}}\atop {\{\Delta y_k\}}}}
    &\sum_{k=0}^{N_p} (\hat{y}_{i,k} +  \Delta y_{k} - y_{k}^{r})^T Q_{k}(\hat{y}_{i,k} + \Delta y_{k} - y_{k}^{r})\\ 
    &+ \sum_{k=0}^{N_p-1}(\hat{u}_{i,k}+\Delta u_k)^TR(\hat{u}_{i,k}+\Delta u_k)\\
    s.t.& \Delta z_{k+1}  = 
    \hat{A}_{i,k}\Delta z_{k} + \hat{B}_{i,k}\Delta u_{k} + A_{i,k}\hat{z}_{i,k}\\
    & \qquad \qquad + \hat{B}_{i,k}\hat{u}_{i,k} - \hat{z}_{i,k+1}, \qquad k=0,\dots,N_p-1\\
    &\hat{y}_{i,k} = C\hat{z}_{i,k},\Delta y_{k} = C\Delta z_{k}, \qquad \quad k=0,\dots,N_p\\
    &E_k(\hat{y}_k\! +\! \Delta y_k)\! +\! F_k(\hat{u}_k\! +\!  \Delta u_k) \!\leq\! l_k, \; k=0,\dots,N_p-1\\
    & E_{N_P} (\hat{y}_{N_P}+\Delta y_{N_P}) \leq l_{N_P}, \hat{z}_{i,0}=\psi_x(x_0).\\
    \end{aligned}  
\end{equation}
Then, the linearized MPC problem is transformed into a dense QP problem and solved efficiently. After the solution is obtained, by adding the increment, we update $\{\hat{z}_k\}, \{\hat{u}_k\}$ at each iteration until the solution is converged
\begin{equation}
    \label{update z_hat, u_hat}
    \begin{aligned}
    \hat{u}_{i+1,k} &  = \hat{u}_{i,k} + \Delta u^*_{i,k},\\
    \hat{z}_{i+1,k} & = \hat{z}_{i,k} + \Delta z^*_{i,k},\\    
    \end{aligned}
\end{equation}
where $\{\Delta z^*_{i,k}\}$ and $\{\Delta u^*_{i,k}\}$ are the optimal solutions of \eqref{I-MPC}.
Besides the estimation update, a initial guess for the first iteration is needed.
The initial guess of $\{\hat{z}_k\}$ and $\{\hat{u}_k\}$ for next time-step is derived from the solution of the current time-step
\begin{equation}
    \label{init z_hat, u_hat}
    \begin{aligned}
    &\hat{u}_{k}  = u^*_{k+1}, k=0,\dots,N_p-2,\\
    &\hat{z}_{k}  = z^*_{k+1}, k=0,\dots,N_p-1,\\    
    &\hat{z}_{N_p} = z^*_{N_p},\hat{u}_{N_p-1} = u^*_{N_p-1}.\\
    \end{aligned}
\end{equation}

\subsection{K-BMPC algorithm}

We summarize the K-BMPC algorithm as follows. At each time-step, the reference $\{ y_k^{r}\}$ is updated. Then we linearize the bilinear MPC problem around the initial guess, solve the linearized problem by applying the iterative QP method. The first step of optimal control is deployed to the system and the initial guess in the next time-step is determined from \eqref{init z_hat, u_hat}.

\begin{algorithm}
\caption{Koopman BMPC}\label{Koopman BMPC}
\begin{algorithmic}
\Statex \textbf{Input:} Initial state $x_0$ and input $u_0$.
\State $\hat{u} \gets u_0 \otimes \textbf{1}_{N_p}, \hat{z} \gets \psi_x(x_0), \otimes \textbf{1}_{N_p+1}$
\While{MPC is running}
    \State Update the reference $\{ y_k^{r}\}, iter \gets 0 $
    \While {Not converged and $iter \leq iter_{max}$}
        \State Form $ \hat{A}_{i,k}, \hat{B}_{i,k}$ from \eqref{A_hat, B_hat}
        \State Solve \eqref{I-MPC} to get the increment $ \Delta u_i$
        \State Update $\{\hat{z}_k\}$ and $\{\hat{u}_k\}$ from \eqref{update z_hat, u_hat}
        \State $iter \gets iter + 1 $
    \EndWhile
    \State Deploy the first control input of $\hat{u}$ to \eqref{dynamic model of tractor-trailer} 
    \State Determine the initial guess of $\{\hat{z}_k\}$ and $\{\hat{u}_k\}$ from \eqref{init z_hat, u_hat}
\EndWhile
\end{algorithmic}
\end{algorithm}
\section{SIMULATION AND EXPERIMENTS RESULTS}
\subsection{Open-loop prediction}
We test the performance of open-loop prediction in simulation. Basic parametric settings of the tractor-trailer system are listed in Table \ref{Parametric settings}.
\begin{table}[htbp]
\caption{Parametric settings for tractor-trailer in simulation}
\begin{center}
\begin{tabular}{c|c|c|c}
\hline
Parameter & Settings & Parameter & Settings\\
\hline
$l_0$ & $3.6m$ & $\varphi_{max}$& $0.6rad$\\
\hline
$l_H$ & $1m$ & $v_{max}$ & $1m/s$\\
\hline
$l_1$ & $6m$ & $a_{max}$ & $2m/s^2$\\
\hline
$\delta\theta_{max}$ & $\pi/3$ & $\omega_{max}$ & $2rad/s$\\
\hline
$\mu$ & $\sim U[0.97, 0.99]$ & $\kappa$ & 0.94\\
\hline
\end{tabular}
\label{Parametric settings}
\end{center}
\end{table}
To collect data for identification, we generate a set of 50000 trajectories, each trajectory is simulated forward 40 steps using 4-th Runge-Kutta method with sampling time $T_s = 0.05s$.

A discrete-time Koopman bilinear model is approximated as introduced in Section III-B, the max derivative order is set to $\rho = 2$. More specifically, we compute the derivatives \eqref{derivative of Fn} under the assumption that $\mu, \kappa = 1$. 
The derivatives of $v$ and $\tan\varphi$ in \eqref{dynamic model of tractor-trailer} are linear to control inputs. Thus we evaluated the errors of position orientation angle $e_{x_0,y_0}$,  $e_{x_1,y_1}$, $e_{\theta_0}$, $e_{\theta_1}$ respectively. As comparison, we consider the following commonly used models:
\begin{itemize}
    \item KBM: Koopman bilinear model.
    \item LKBM: locally linearized Koopman bilinear model at $z_0$, which is used in \cite{Bruder 2021}.
    \item NM: the nominal model of \eqref{dynamic model of tractor-trailer}, considering $\mu, \kappa=1$.
    \item LLNM: locally linearized nominal model at $x_0$ and $u_0$.  
\end{itemize}

\begin{figure}[htbp]
\centerline{\includegraphics[width=0.7\linewidth]{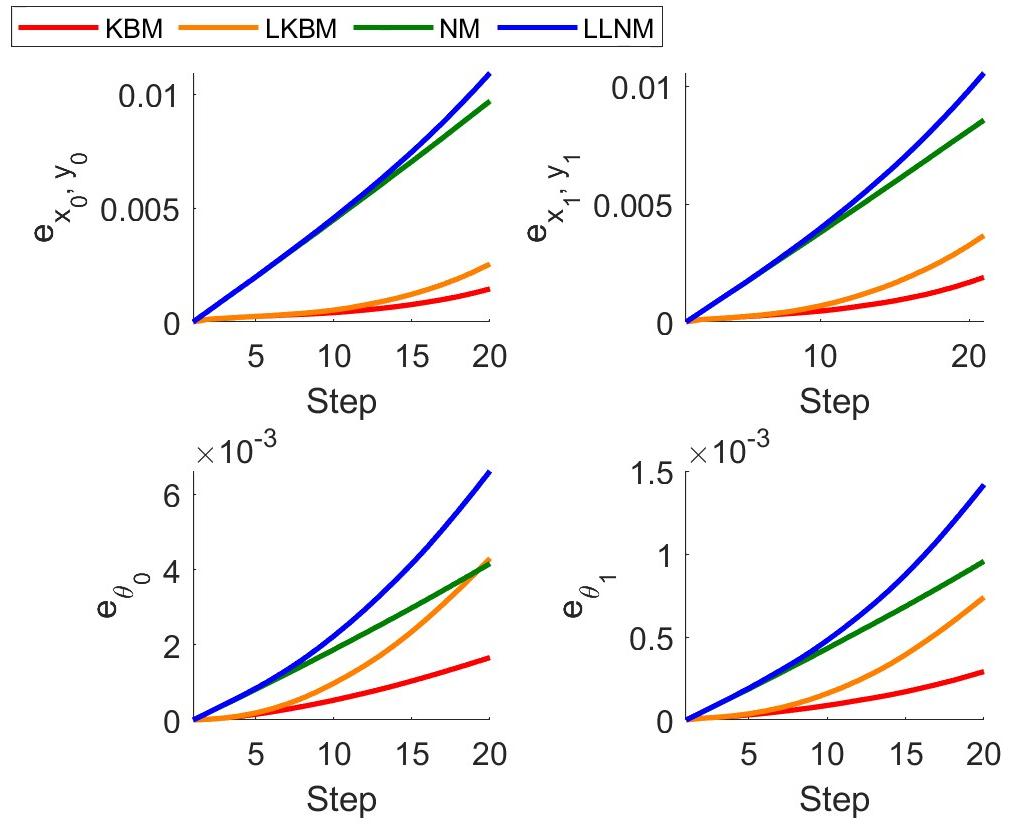}}
\caption{Propagation of state prediction errors for different models.}
\label{prediction error}
\end{figure}

We test the prediction performance of the above models under 1000 randomized initial states and control sequences, each initial state is simulated forward 20 steps. The errors (denoted as $e_{x_0,y_0}(10^{-4}m), e_{x_1, y_1}(10^{-4}m), e_{\theta_0}(10^{-4}rad)$ and $e_{\theta_1}(10^{-4}rad)$) are averaged to measure the errors between the truth and the prediction. 
Table \ref{Mean prediction errors of Koopman bilinear model and locally linearized model under randomized control inputs} presents the mean prediction errors and Fig. \ref{prediction error} depicts propagation of state prediction errors for each model.
\begin{table}[htbp]
\caption{Mean prediction errors under randomized control inputs}
\begin{center}
\begin{tabular}{c|c|c|c|c}
\hline
 & $e_{x_0,y_0}$& $e_{x_1,y_1}$& $e_{\theta_0}$ & $e_{\theta_1}$\\
\hline
KBM & \textbf{6.12}& \textbf{6.94}& \textbf{7.18}& \textbf{1.24}\\
\hline
LKBM & $9.48$& $12.18$& $16.35 $& $2.78$\\
\hline
NM & $50.44$ & $42.45$& $21.31$& $4.90$\\
\hline
LLNM& $53.90$ & $47.63$& $29.57$& $6.33$\\
\hline
\end{tabular}
\label{Mean prediction errors of Koopman bilinear model and locally linearized model under randomized control inputs}
\end{center}
\end{table}

As can be seen from Fig.\ref{prediction error}, Koopman bilinear model exhibits superior prediction performance over NM and LLNM. The performance improvements can be attributed to the identification of slip effects through the Koopman bilinear model. On the other hand, LKBM does not create noticeable discrepancies from the KBM and aligns closely with the ground truth. Nevertheless, as we shall see the linearization at $z_0$ brings extra prediction errors, we linearize KBM around $\hat{z}$ and $\hat{u}$ to reduce prediction errors. 
\subsection{Closed-loop tracking control}

\subsubsection{Simulation results}
We implement our algorithm in Matlab 2022b. We select the predictive horizon as $N_p = 20$, relevant controller parameters are assigned as follows: $Q = diag[10,10,1,1,0,0,10,10]$, $R = diag[0.01, 1]$, $Q_{N_p} = 10Q$, $iter_{max} = 3$. 
Nominal nonlinear MPC (NMPC)  and iteratively locally linearized nominal MPC (LMPC) are chosen as benchmarks. NMPC uses the nominal model as predictive model, LMPC solves the linearized NMPC problem iteratively in a similay way to section IV-C. All simulations are run with the same settings.

\begin{figure}[htbp]
\centering
\subfigure[Tractor-trailer tracking trajectory results in simulation]{\includegraphics[width=0.35\hsize]{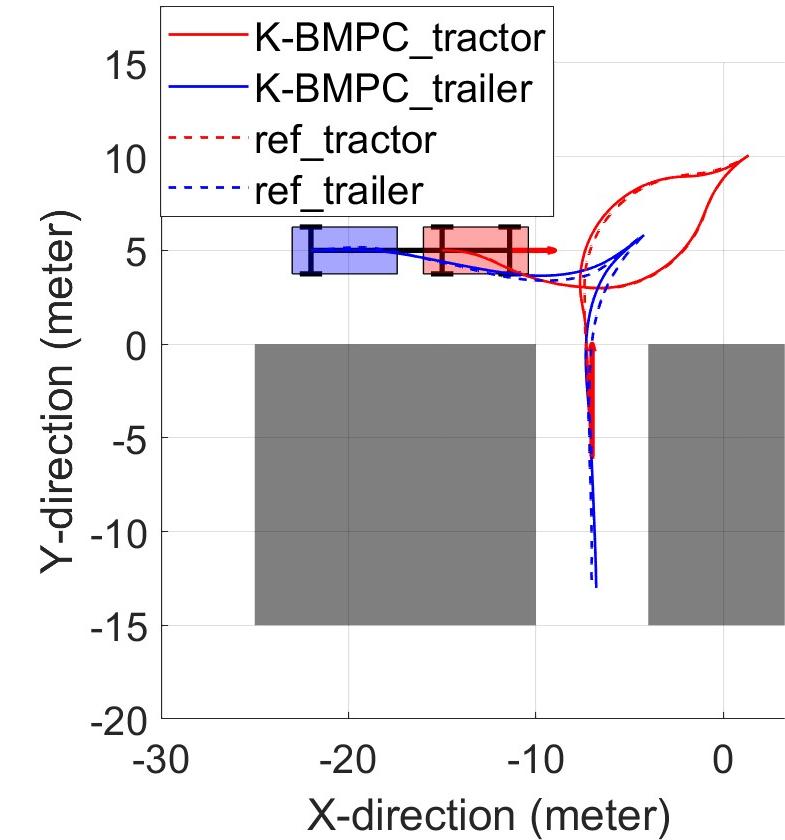}
\label{simulation traj tracking}}
\hspace{0cm}
\subfigure[Comparison of tracking trajectory results in simulations and experiments.]
{\includegraphics[width=0.55\hsize]{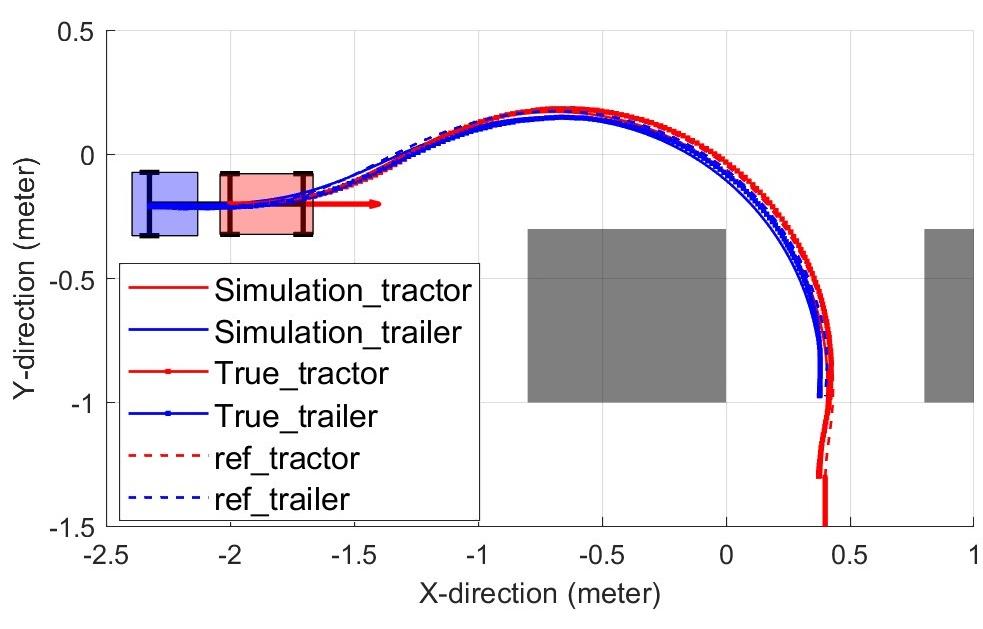}
\label{simulation and real traj tracking}}
\caption{Tractor-trailer trajectory when tracking a given path}
\label{Tractor-trailer trajectory when tracking a given path}
\end{figure}

\begin{figure}[htbp]
\centering
\subfigure[Tracking errors for all time-steps.]{\includegraphics[width=0.45\hsize]{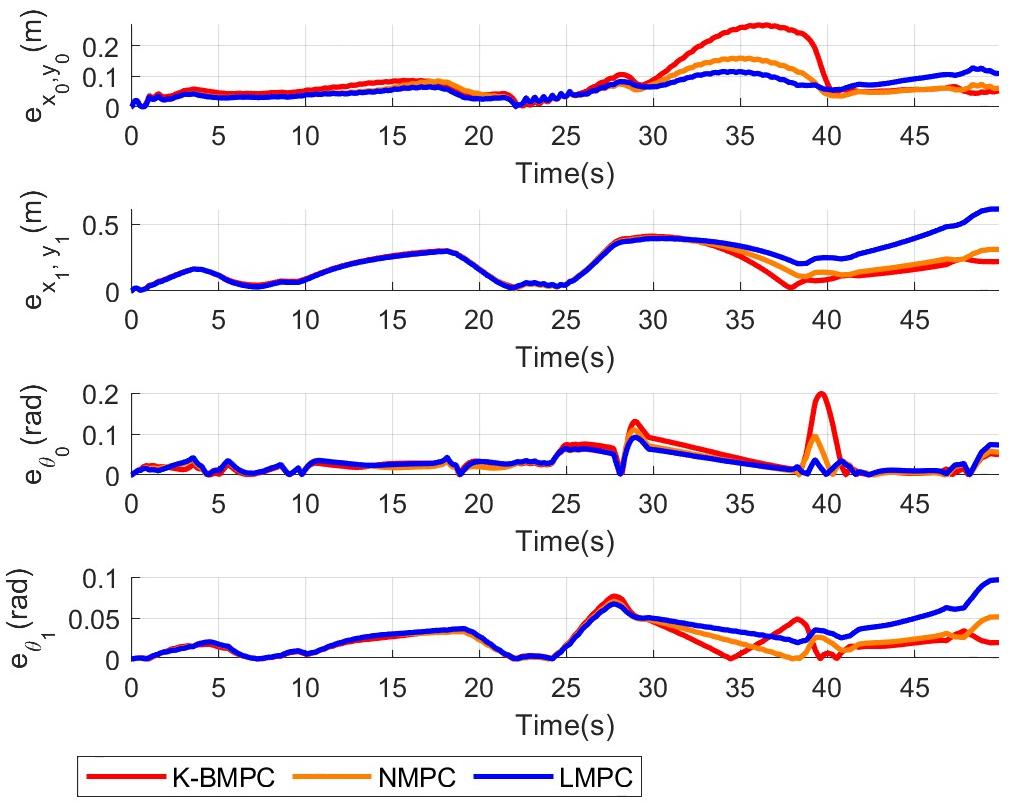}
\label{Tracking errors and control inputs}}
\hspace{0cm}
\subfigure[Average cost and solving time for all time-steps.]
{\includegraphics[width=0.45\hsize]{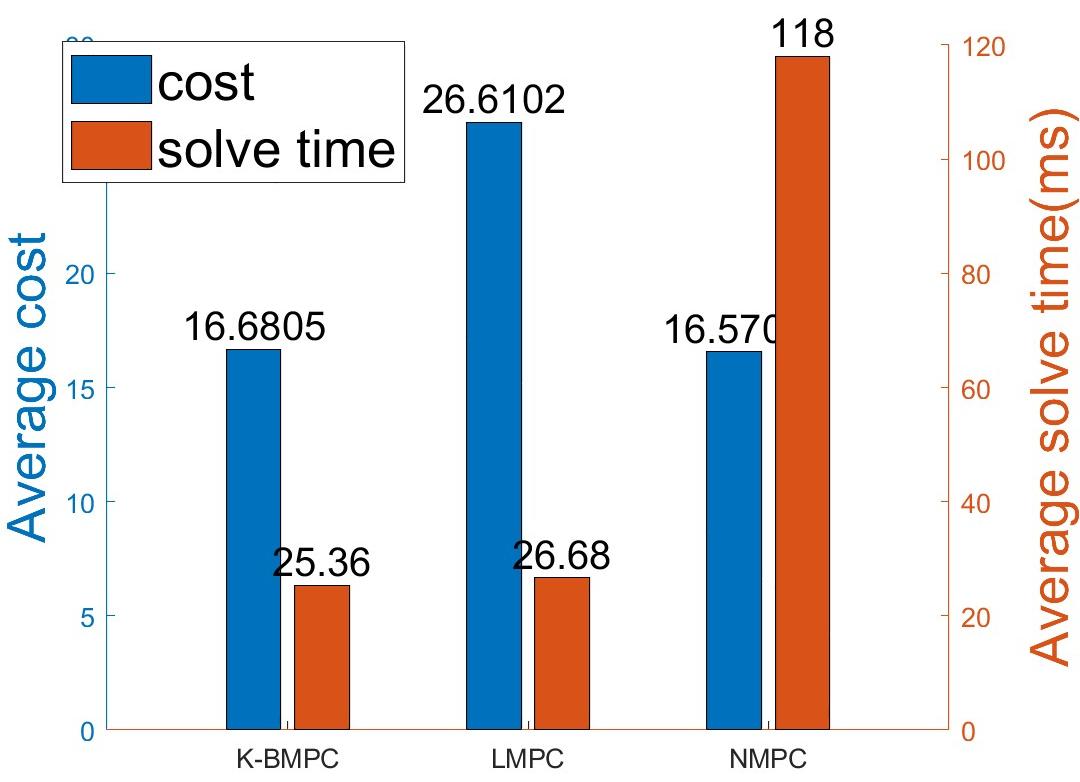}
\label{cost compute time}}
\caption{Tracking results comparison between different methods in simulation.}
\label{Tracking results in simulation.}
\end{figure}

In simulation, the tractor-trailer needs to follow a reference trajectory and park vertically. To this end, the reference trajectory is given by the optimization-based method\cite{Li 2021} with the discretization time $T \approx 0.5s$,  thereby failing to satisfy the dynamics in a small sampling time $T=0.05s$. 
The result trajectory is depicted in Fig. \ref{simulation traj tracking}, and the trajectory tracking errors $e_{x_0,y_0}, e_{x_1, y_1}, e_{\theta_0}$, $e_{\theta_1}$ are shown in Fig. \ref{Tracking errors and control inputs}. The average MPC cost and solving time for all time-steps are presented in Figure \ref{cost compute time}, mean tracking errors of the controllers are displayed in Table \ref{Mean Error of trajectory tracking}. 
\begin{table}[htbp]
\caption{Mean tracking errors comparison between different methods}
\begin{center}
\begin{tabular}{c|c|c|c|c}
\hline
 & $e_{x_0,y_0}(m)$& $e_{x_1,y_1}(m)$& $e_{\theta_0}(rad)$ & $e_{\theta_1}(rad)$\\
\hline
K-BMPC& 0.0861 & \textbf{0.1783} & 0.0347& \textbf{0.0213}\\
\hline
NMPC& 0.0627& 0.1903& \textbf{0.0281}& 0.0223\\
\hline
LMPC& \textbf{0.614}& 0.2389& 0.0283& 0.0303\\
\end{tabular}
\label{Mean Error of trajectory tracking}
\end{center}
\end{table}

From Table \ref{Mean Error of trajectory tracking}, we find that the mean trailer position and orientation tracking errors of K-BMPC are lower than that of NMPC and LMPC, while NMPC has lower errors in tractor orientation tracking, LMPC has lower tractor position error. 
Overall, K-BMPC demonstrates superior tracking performance when compared to LMPC. The improvement in tracking accuracy is due to the fact that Koopman bilinear model identifies slip parameters from data and predicts the evolution of high order derivatives in tractor-trailer dynamics. 
Finally, the data from Figure \ref{cost compute time} indicates that LMPC has a cost 59.53\% higher than K-BMPC, while the cost of NMPC is sightly lower. Moreover, the average solving time for K-BMPC is significantly less than that of NMPC. 
\subsubsection{Experimental results}
\begin{figure}[htbp]
\centerline{\includegraphics[width=0.7\linewidth]{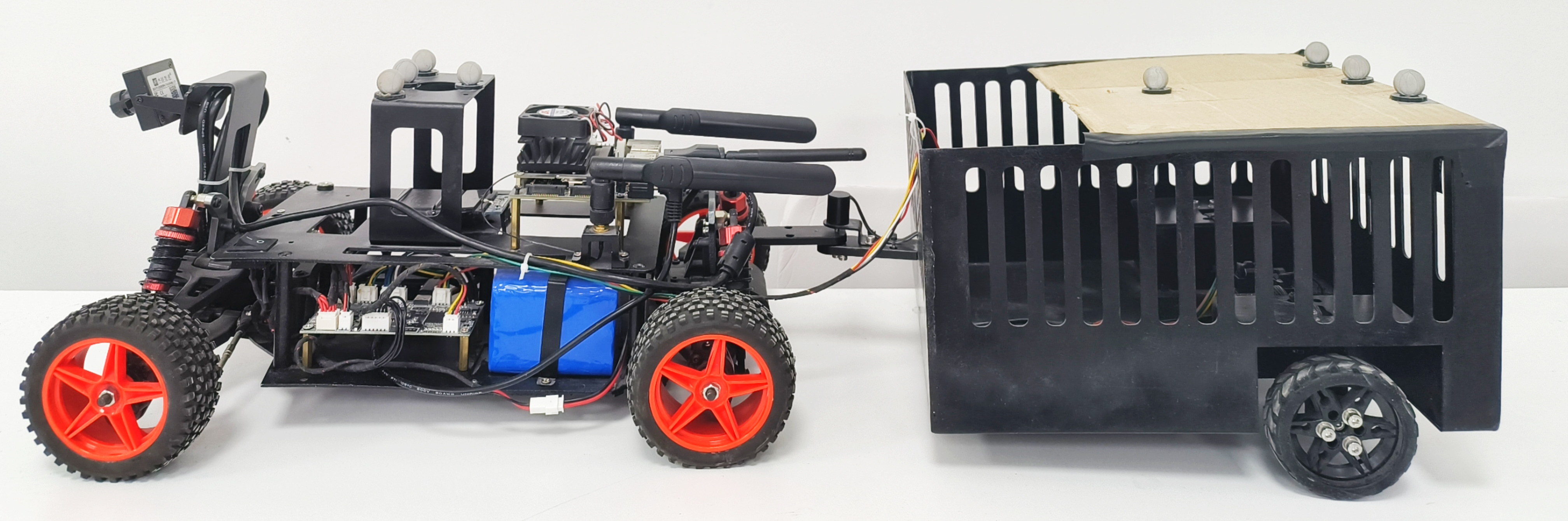}}
\caption{Miniature tractor-trailer system.}
\label{real tractor-trailer}
\end{figure}

Furthermore, we test our algorithm on a miniature tractor-trailer model, as shown in Fig. \ref{real tractor-trailer}.
We use Vicon motion capture system to measure the global positions of the tractor and trailer and a virtual machine with 8GB RAM to compute control inputs. The tractor-trailer model receives and executives the control inputs from the virtual machine.
Communication between the Vicon, virtual machine and the tractor-trailer miniature model is achieved by Robot Operating System (ROS). 
Fig. \ref{simulation and real traj tracking} and Table \ref{real and sim atrajectory tracking error} detail the tracking performance of the K-BMPC algorithm on a 20-second trajectory in simulation and real-world experiment.
\begin{table}[htbp]
\caption{Mean tracking errors in simulations and experiments}
\begin{center}
\begin{tabular}{c|c|c|c|c}
\hline
 & $e_{x_0,y_0}(m)$& $e_{x_1,y_1}(m)$& $e_{\theta_0}(rad)$ & $e_{\theta_1}(rad)$\\
\hline
Real model & 0.0378& 0.0420& 0.0258& 0.0521\\
\hline
Simulation & 0.0284& 0.0297& 0.0094& 0.0089\\
\hline
\end{tabular}
\label{real and sim atrajectory tracking error}
\end{center}
\end{table}

The results show that while the average tracking errors in the real-model application are higher than that in the simulation, K-BMPC is still able to control the tractor-trailer to track the reference trajectory. The increased errors are likely attributed to factors such as the initial condition errors, topic transmission delay and the uncertainty and dead zone of the actuator.

\section{CONCLUSION}
This paper presents K-BMPC, a Koopman-based bilinear MPC for tractor-trailer tracking control. We propose a derivative-based lifting function construction methods to approximate tractor-trailer dynamics with unknown parameters using a Koopman bilinear model. Moreover, when the derivative order is truncated, we analyze the state prediction error propagation and its bounds. K-BMPC linearizes the bilinear term around the estimation of lifted state and control input then solves Koopman bilinear MPC problems iteratively. Closed-loop tracking results show the proposed K-BMPC exhibits elevated tracking precision along with commendable computational efficiency. Compared to LMPC, K-BMPC keeps better tracking accuracy with a similar solving time. The real-world experiments help to verify the feasibility of the method. 
Future work will be devoted to identify a Koopman bilinear model for a tractor-trailer vehicle using experimental data, along with analyzing the robustness of Koopman bilinear MPC under the localization and modeling uncertainties.

\end{document}